\documentstyle[preprint,revtex]{aps}

\begin{document}
\preprint{IFUSP/P-974/1992}
\draft

\begin{title}
{Coherent States of the $SU(N)$ groups}
\end{title}

\author
{D.M. Gitman}
\begin{instit}
Instituto de F\'{\i}sica \\
Dept. de F\'{\i}sica Matem\'atica,\\ Universidade de S\~ao Paulo,\\
Caixa Postal 20516\\
01498 - S\~ao Paulo, S.P., Brazil.
\end{instit}

\author{A.L. Shelepin}
\begin{instit}
Moscow Institute of Radio Engineering,\\ Electronics and
Automation, Prospect\\ Vernadskovo 78, Moscow 117454, \\Russia.
\end{instit}

\begin{abstract}

Coherent states $(CS)$ of the $SU(N)$ groups are constructed explicitly and
their properties are investigated. They represent a nontrivial generalization
of
the spining $CS$ of the $SU(2)$ group. The $CS$ are parametrized by the points
of  the coset space, which is, in that particular case, the projective space
$CP^{N-1}$ and plays the role of the phase space of a corresponding classical
mechanics. The $CS$ possess of a minimum uncertainty, they minimize
an invariant dispersion of the quadratic Casimir operator. The classical limit
is ivestigated in terms of symbols of operators. The role of the Planck
constant playes $h=P^{-1}$,  where $P$ is the signature of the representation.
The classical limit of the so called star commutator generates the  Poisson
bracket in the $CP^{N-1}$ phase space. The logarithm of the modulus of the $CS$
overlapping, being interpreted as a symmetric in the space, gives the
Fubini-Study metric in $CP^{N-1}$. The $CS$  constructed are useful for the
quasi-classical analysis of the quantum equations of the $SU(N)$ gauge
symmetric
theories.

\end{abstract}
\newpage

\section{Introduction}

As known, coherent states $(CS)$ are widely and fruitful being utilized in
different areas of the theoretical physics \cite{ab1}-\cite{ab5}. The $CS$,
introduced
by
Schr\"odinger and Glauber, turned out to be orbits of the Heisenberg-Weyl
group.
That observation allowed one to formulate by analogy some general definition of
$CS$ for any Lie group \cite{ab4} . A connection between the $CS$ and the
quantization of classical systems, in particular , systems with a curved phase
phase space, was also established \cite{ab6,ab7} . By the origin, $CS$ are
quantum states, but, at the same time, they are parametrized by the points of
the phase space of a corresponding classical mechanics. Namely that
circumstance
makes
them
very convenient in the analysis of problems of the correspondance between the
quantum and the classical discription. All that explains the interest both to
the investigation of general problems of $CS$ theory and to the construction of
$CS$ of concrete groups. The $CS$ of such important in physics groups as
$SU(N)$ ones
are built and investigated in an uniform way in the present work. The $CS$ of
the
group $SU(2)$, from that family, are well known. One can point out some of the
first references \cite{ab8}-\cite{ab12}, where that states were built on the
base of the
well investigated structure of the $SU(2)$ matrices in the fundamental
representation.  Another approach to
the $CS$ construction of the $SU(2)$ group was used by Berezin \cite{ab6,ab7} .
That approach is connected with the utilization of the representations of the
$SU(2)$ group in the space of  polynomials of the powers not more that a
given one. A modification of the latter method in a gauge invariant form (with
extended number of variables in the coset space or phase space) allows us to
build the $CS$ for all the groups $SU(N)$ in an uniform way.
We construct the $CS$
by means of  orbits of highest weights in the space of polynomials of a
fixed power. The representations used are equivalent to the total symmetric
irredicible unitary representations of the $SU(N)$ groups. The stationary
subgroups of the highest weights, in the case of consideration, are $U(N-1)$,
so
that the $CS$ are parametrized by the points of the coset space $SU(N)/U(N-1)$
which plays the role of the phase space of a corresponding classical mechanics,
and at the same time it is the well known projective space $CP^{N-1}$. The
logarithm of the modulus of the $CS$ overlapping, being interpreted as a
symmetric in
the space $CP^{N-1}$, generates Fubini-Study metric in the space.The $CS$ form
an overcomplited system in the representation space and minimize an
invariant dispersion of the quadratic Casimir operator.
The classical limit is investigated in terms of the operators symbols which are
constructed as the mean values in the $CS$. The role of the Planck constant
plays the quantity $h=P^{-1}$, where $P$  is the signature of the
representation. The
classical limit of so called star commutator of symbols generates the classical
Poisson bracket in the coresponding phase space. The present work is the
continuation of our papers \cite{ab13}-\cite{ab15}, where a part of results was
 preliminary expounded.

\section{The construction of the $CS$ by means of the representations
of the $SU(N)$ groups on polynomials}

We are going to construct $CS$ of the $SU(N)$ groups as orbits in some
irreducible representations of the groups, factorized with respect to
stationary subgroups. First,  we discribe the corresponding representations.

Let ${\bf C}^N$    is $N$-dimentional space of complex lines
$z=(z_\mu),\mu=\overline{1,N}$, with the scalar
product $(z,z')_C=\sum_{\mu}\bar{z}_\mu z'_\mu,\: \mu=\overline{1,N}$,
and $\widetilde{\bf C}^N$    is the dual space of complex columns   with the
scalar product
$(\tilde{z},\tilde{z}')_{\widetilde{C}}=\sum_{\mu}\bar{\tilde{z}^\mu}\tilde{z}^
\mu $. The anti-isomorphism is given by the relation
$z\leftrightarrow\tilde{z}\Leftrightarrow\bar{z}_\mu=\tilde{z}^\mu $.
The mixed (Dirac) scalar product between elements of
${\bf C}^N$    and $\widetilde{\bf C}^N$      is defined by the equation:
\begin{equation} \label{e1}
\langle z',\tilde{z}\rangle =(\tilde{z}',\tilde{z})_{\widetilde{C}}=
\overline{(z',z)_C}=z'_\mu\tilde{z}^\mu\:.
\end{equation}

Let $g$ are matrices of the fundamental representation of the $SU(N)$ group.
This representation induces irreducible representations of the group in the
spaces $\Pi_P$ and    $\widetilde{\Pi}_P$ of polynomials of a fixed power $P$
on the  vectors $z$  and $\tilde{z}$  respectively, \begin{eqnarray}
T(g)\Psi_P(z)&=&\Psi_P(z_g),\: z_g=zg,\: \Psi_P\in\Pi_P\:,\nonumber\\
\widetilde{T}(g)\Psi_P(\tilde{z})&=&\Psi_P(\tilde{z}_g),\: \tilde{z}_g=g^{-1}
\tilde{z},\:\Psi_P(\tilde{z})\in\widetilde{\Pi}_P\:.\label{e2} \end{eqnarray}

\noindent The anti-isomorphism $z\leftrightarrow\tilde{z}$ induces the
correspondence $\Psi_P(\tilde{z})=\overline{\Psi_P(z)}$.

The representation (\ref{e2}) is equivalent to the one on total symmetrical
tensors of
signature $P$ . So, we will further call $P$ as the signature of the
irreducible
representation.

Obviously the monomials
\begin{eqnarray}
\Psi_{P,\{n\}}(z)=\sqrt{\frac{P!}{n_1!...n_N!}}z_1^{n_1}...z_N^{n_N}\:,\label
{e3}\\
 \{n\}=\{n_{1},\ldots,n_{N}|\sum_{\mu}n_{\mu}=P\}\:,\nonumber
\end{eqnarray}

\noindent form a discrete basis in $\Pi_{P}$ , and the monomials $\Psi_{P,
\{n\}}(\tilde{z})=\overline{\Psi_{P,\{n\}}(z)}$   form a basis in
$\widetilde{\Pi_{P}}$. The monomials obey the remarkable relation
\begin{equation}\label{e4}
\sum_{\{n\}}\Psi_{P,\{n\}}(z')\Psi_{P,\{n\}}(\tilde{z})=\langle
z',\tilde{z}\rangle^{P}\:,
\end{equation}

\noindent which is group  invariant on account of the invariance  of the
scalar product (\ref{e1}) under the group transformation, $\langle
z'_{g},\tilde
{z}_{g}\rangle=\langle z',\tilde{z}\rangle$ .

We  introduce  also  the scalar product of two polynomials:
\begin{eqnarray}\label{e5}
\langle\Psi_{P}|\Psi'_{P}\rangle&=&\int\overline{\Psi_{P}(z)}\Psi'_{P}
(z){\rm d}\mu_{P}(\bar{z},z)\:,\\
{\rm d}\mu_{P}(\bar{z},z)&=&\frac{(P+N-1)!}{(2\pi)^{N}P!}\delta(\sum|
z_{\mu}|^{2}-1)\prod{\rm d}\bar{z}_{\nu}{\rm d}z_{\nu}\:,\nonumber\\
{\rm d}\bar{z}{\rm d}z&=&{\rm d}(|z|^{2}){\rm d}(\arg z)\:\nonumber.
\end{eqnarray}

\noindent Using the integral

\[
\int_{0}^{1}{\rm d}\rho_{1}\ldots\int_{0}^{1}{\rm d}\rho_{N}\delta(\sum\rho_
{\mu}-1)\prod_{\nu=1}^{N}\rho_{\nu}^{n_{\nu}}=\frac{\prod_{\nu=1}^{N}n_{\nu}!}
{\left(\sum_{\nu=1}^{N}n_{\nu}+N-1\right)!}\:,
\]

\noindent it is easy to verify that the orthonormality relation holds:
\begin{equation}\label{e6}
\langle\Psi_{P,\{n\}}|\Psi_{P,\{n'\}}\rangle=\langle P,n|P,n'\rangle=
\delta_{\{n\},\{n'\}}\:.
\end{equation}

\noindent The copmpleteness relation take place as well
\begin{equation}\label{e7}
\sum_{\{n\}}|P,n\rangle\langle P,n|=I_{P}\:,
\end{equation}
\noindent
where $|P,n\rangle$ and $\langle P,n|$ are Dirac's denotations for the vectors
$\Psi_{P,\{n\}}(z)$ and $\Psi_{P,\{n\}}(\tilde{z})$
respectively, and $I_{P}$ is the identical operator in the irreducible space
of representation of signature $P$ .

It is covenient to introduce the operators $a_{\mu}^{\dag}$ and $a^{\mu}$
which act on the basis vectors by formulas:
\begin{eqnarray}
a_{\mu}^{\dag}\Psi_{P,\{n\}}(z)&=&z_{\mu}\Psi_{P,\{n\}}(z)\:\rightarrow
\:a_{\mu}^{\dag}|P,n\rangle=\sqrt{\frac{n_{\mu}+1}{P+1}}|P+1,\ldots,n_
{\mu}+1,\ldots\rangle\:,\nonumber\\
a^{\mu}\Psi_{P,\{n\}}(z)&=&\frac{\partial}{\partial z_{\mu}}\Psi_{P,\{n\}}
(z)\:\rightarrow\:a^{\mu}|P,n\rangle =\sqrt{Pn_{\mu}}
|P-1,\ldots,n_{\mu}-1,\ldots\rangle\:,\nonumber\\
{}[a^{\mu},a_{\nu}^{\dag}]&=&\delta^{\mu}_{\nu},\;
[a^{\mu},a^{\nu}]=[a^{\dag}_
{\mu},a_{\nu}^{\dag}]=0\:.\label{e8}
\end{eqnarray}

\noindent One can find the action of these operators on the left,

\begin{eqnarray}
\langle P,n|a_{\mu}^{\dag}&=&\sqrt{\frac{n_{\mu}}{P}}\langle P-1,\ldots,n_{\mu}
-1,\ldots|=\frac{1}{P}\frac{\partial}{\partial\tilde{z}^{\mu}}\Psi_{P,\{n\}}(
\tilde{z})\:,\label{e9}\\
\langle P,n|a^{\mu}&=&\sqrt{(P+1)(n_{\mu}+1)}\langle P+1,\ldots,n_{\mu}+1,
\ldots|=(P+1)\tilde{z}^{\mu}\Psi_{P,\{n\}}(\tilde{z})\:.\nonumber
\end{eqnarray}

\noindent Their quadratic combinations $A^{\nu}_{\mu}$  can serve as generators
in each irreducible representation of signature $P$ ,

\begin{eqnarray}
A^{\nu}_{\mu}&=&a_{\mu}^{\dag}a^{\nu}=z_{\mu}\frac{\partial}{\partial z_{\nu}},
\;  \left[A^{\nu}_{\mu},A^{\kappa}_{\lambda}\right]=\delta^{\nu}_{\lambda}A^
{\kappa}_{\mu}-\delta^{\kappa}_{\mu}A^{\nu}_{\lambda}\:,\label{e10}\\
A^{\nu}_{\mu}|P,n\rangle&=&\sqrt{n_{\nu}(n_{\mu}+1)}|P,\ldots,n_{\nu}-1,\ldots,
n_{\mu}+1,\ldots\rangle\:,\;\mu\neq\nu\:,\nonumber\\
A_{\mu}^{\mu}|P,n\rangle&=&n_{\mu}|P,n\rangle,\;  \sum_{\mu}A^{\mu}_{\mu}|P,n
\rangle=P|P,n\rangle\:.\nonumber
\end{eqnarray}

\noindent Obviously, the $A^{\mu}_{\mu}$ are Cartan's generators and $(n_{1},
\ldots,n_{N})$ the weight vector.

The independent generators $\hat{\Gamma}_{a},\:a=\overline{1,N^{2}-1},$
can be expressed in terms of the operators $A_{\mu}^{\nu}$:

\begin{equation}\label{e11}
\hat{\Gamma}_{a}=\left(\Gamma_{a}\right)^{\nu}_{\mu}A^{\mu}_{\nu},\;
\left[\hat{\Gamma}_{a},\hat{\Gamma}_{b}\right]=\imath f_{abc}\hat{\Gamma}_{c}
\;,
\end{equation}
where $\Gamma_{a}$ are generators in the fundamental representation, $\left[
\Gamma_{a},\Gamma_{b}\right]=\imath f_{abc}\Gamma_{c}$.

The quadratic Casimir operator $C_{2}=\sum_{a}\hat{\Gamma}_{a}^{2}$ can be only
expressed via the operators $A^{\nu}_{\mu}$ by means of the well known formula
\begin{equation}\label{e12}
\sum_{a}\left(\Gamma_{a}\right)^{\nu}_{\mu}\left(\Gamma\right)^{\kappa}_
{\lambda}=\frac{1}{2}\delta^{\nu}_{\lambda}\delta^{\kappa}_{\mu}-\frac{1}
{2N}\delta^{\nu}_{\mu}\delta^{\kappa}_{\lambda}\;,
\end{equation}
and evaluated in every irreducible representation explicitly,

\begin{equation}\label{e13}
C_{2}=\frac{1}{2}\tilde{A}^{\nu}_{\mu}\tilde{A}^{\mu}_{\nu}=\frac{P(N+P)(N-1)}
{2N},\;     \tilde{A}^{\nu}_{\mu}-\frac{\delta_{\mu}^{\nu}}{N}\sum_{\lambda}A^
{\lambda}_{\lambda}\;.
\end{equation}

Now we are goig to construct the orbits of highest weights ( of a vector of the
basis (\ref{e3}) with the maximal length $\sqrt{\sum n^{2}_{\mu}}=P$ ). Let
this highest weight be the state $\Psi_{P,\{P,0\ldots 0\}}(z)=(z_{1})^{P}$.
Then we get,
in accordance with (\ref{e2}) :
\begin{equation}\label{e14}
T(g)\Psi_{P,\{P,0\ldots \}}(z)=\left[z_{\mu}g^{\mu}_{1}\right]^{P}=\langle
z,\tilde{u}\rangle^{P},\;   \tilde{u}^{\mu}=g^{\mu}_{1}\;,
\end{equation}

\noindent where the vector $\tilde{u}\in\widetilde{\bf C}^{N}$ is the first
column of the $SU(N)$  matrix in the fundamental representation.

If we interprete the representation space as a Hilbert one of quantum states,
then we have to identify all the states differ each other by a constant phase.
Let us turn from that point of view to the states of the orbit (14). One can
notice, that the transformation  $\arg\tilde{u}^{\mu}\rightarrow\arg\tilde{u}^
{\mu}+\lambda$,  changes  all the states (\ref{e14}) by the constant phase
$\exp(iP\lambda)$.
So, one can treate the transformation as gauge one in certain sense. To select
only physical different quantum states $(CS)$ from all the states of the orbit,
one has to impose a gauge condition on $\tilde{u}$ which fixes the total phase
of the orbit (\ref{e14}) . Such a condition may be chosen in the form $\sum_{
\mu}\arg\tilde{u}^{\mu}=0$.

Taken into account that the quantities $\tilde{u}$ obey the condition
$\sum|\tilde{u}^{\mu}|^{2}=1$,  by the origin,  as elements of the first column
of the $SU(N)$ matrix, we get the explicit form of the $CS$ of the $SU(N)$
group in the space $\Pi_{P}$:

\begin{eqnarray}
\Psi_{P,\tilde{u}}(z)=\langle z,\tilde{u}\rangle^{P}\;,\label{e15}\\
\sum_{\mu}|\tilde{u}^{\mu}|^{2}=1,\;  \sum_{\mu}\arg \tilde{u}^{\mu}=0 .
\label{e16}
\end{eqnarray}

\noindent In the same manner we construct the orbit of the highest weight $\Psi
_{P,\{P,0\ldots 0\}}(\tilde{z})=\left(\tilde{z}^{1}\right)^{P}$ in the space
$\tilde{\Pi}_{P}$ ,  and the corresponding $CS$ have the form:

\begin{eqnarray}
\Psi_{P,u}(\tilde{z})=\langle u,\tilde{z}\rangle^{P}, \label{e17}\\
\sum_{\mu}|u_{\mu}|^{2}=1,\;  \sum_{\mu}\arg u_{\mu}=0 .\label{e18}
\end{eqnarray}

\noindent Obviously,  $\Psi_{P,\tilde{u}}(z)=\overline{\Psi_{P,u}(\tilde{z})},
\;z\leftrightarrow\tilde{z},  u\leftrightarrow\tilde{u}$.

It is easy to see that all the elements of the discrete basis (\ref{e3}) with
the weight vectors of the form $(n_{\mu}=\delta^{\nu}_{\mu}P,\: \mu=\overline
{1,N})$
belong to the $CS$ set (\ref{e15}) with parameters $(\tilde{u}^{\mu}=\delta^{
\mu}_{\nu}P,\;\mu=\overline{1,N})$. The analogous statement is valid regarding
to the dual basis and to the $CS$  (\ref{e17}) .

The quantities $\tilde{u}$   and $u$, which parametrize the $CS$  (\ref{e15})
 and (\ref{e16}) ,
are elements of the coset space $SU(N)/U(N-1)$, in accordance with the fact
that the stationary subgroups of both the initial vectors from the spaces $\Pi_
{P}$ and $\tilde{\Pi}_{P}$ are $U(N-1)$. At the same time, the coset space is
the so called projective space $CP^{N-1}$  (we remember that the complex
projective space is defined as a set of all nonzero vectors $z$ in ${\bf C}^{N}
$, where $z$ and $\lambda z,\:\lambda\neq 0$, are equivalent \cite{ab16} ). The
 eq.(\ref{e15}) or (\ref{e17}) , are just the possible conditions which define
the projective space. The coordinates $u$ or $\tilde{u}$ are called
homogeneouse ones
in the $CP^{N-1}$. Thus, the $CS$ constructed are parametrized by the elements
of
the projective space $CP^{N-1}$, which is a symplectic manifold \cite{ab16} and
therefore can be considered as the phase space of a classical mechanics.

To decompose the $CS$ in the discrete basises, one can use the scalar product
(\ref{e5}) directly, but there exists more simple way. One can use the relation
(\ref{e4}) , on
account of the right side of eq. (\ref{e4}) can be treated as $CS$  (\ref{e15})
 or
(\ref{e17}) . So,
it follows from (\ref{e4}) :
\begin{equation}\label{e19}
\Psi_{P,\tilde{u}}(z)=\sum_{\{n\}}\Psi_{P,\{n\}}(\tilde{u})\Psi_{P,\{n\}}(z)\:.
\end{equation}
That implies:

\begin{equation}\label{e20}
\langle P,u|P,n\rangle=\Psi_{P,\{n\}}(u),\:  \langle P,n|P,u\rangle=\Psi_{P,\{n
\}}(\tilde{u}),
\end{equation}
where $|P,u\rangle$  and $\langle P,u|$  are Dirac's denotations for the $CS
 \; \Psi_{P,\tilde{u}}(z)$ and $\Psi_{P,u}(\tilde{z})$ respectively.So we come
to  the important for the understanding result: the discrete basises in the
spaces
$\Pi_{P}$ and $\tilde{\Pi}_{P}$ are ones  in the $CS$ representation.

The completeness relation for the $CS$ can be extracted from the eq.(\ref{e6}).
Using
the formulas (\ref{e20}) in the integral (\ref{e6}) , we get:
\[
\int\langle P,n|P,u\rangle\langle P,u|P,n'\rangle\rm d\mu_{P}(\bar{u},u)=
\delta_{\{n\},\{n'\}}\;.
\]
That proves the completeness relation
\begin{equation}\label{e21}
\int|P,u\rangle\langle P,u|\rm d\mu_{P}(\bar{u},u)=I_{P}\;.
\end{equation}

\section {Uncertainty relation}

The orbit of each vector of the discrete basis $|P,n\rangle$  (\ref{e3}) and ,
 particularly, the $CS$ constructed, are eigen for a nonlinear operator $C'_{2}
$, which is defined by its action on an arbitrary vector $|\Psi\rangle$  as
\begin{equation}\label{e22}
C'_{2}|\Psi\rangle=\sum_{a}\langle\Psi|\hat{\Gamma}_{a}|\Psi\rangle\hat{\Gamma}_
{a}|\Psi\rangle\;.
\end{equation}
First, we note that $T^{-1}(g)C'_{2}T(g)=C'_{2}$ , where $T(g)$  are operators
of the representation. Indeed, it follows from the relation $T^{-1}(g)\hat{
\Gamma}_{a}T(g)=t^{c}_{a}\hat{\Gamma}_{c}$  and $[C_{2},T(g)]=0$, that $t^{c}_
{a}$ is an orthogonal matrix, so that
\begin{eqnarray*}
T^{-1}(g)C'_{2}T(g)|\Psi\rangle&=&\sum_{a}\langle\Psi|T^{-1}(g)\hat{\Gamma}_{a}
T(g)|\Psi\rangle T^{-1}(g)\hat{\Gamma}_{a}T(g)|\Psi\rangle \\
&=&\sum_{a}\langle\Psi|\hat{\Gamma}_{a}|\Psi\rangle\hat{\Gamma}_{a}|\Psi\rangle=
C'_{2}|\Psi\rangle\;.
\end{eqnarray*}
After that, it is easy to show, that the orbit $T(g)|P,n\rangle$ is eigen for
$C'_{2}$ .We write:
\begin{equation}\label{e23}
C'_{2}T(g)|P,n\rangle=T(g)C'_{2}|P,n\rangle=T(g)\sum_{a}\langle
P,n|\hat{\Gamma}
_{a}|P,n\rangle\hat{\Gamma}_{a}|P,n\rangle\:,
\end{equation}

\noindent and use the formulas (\ref{e11}) , (\ref{e12}) in the right side of
(\ref{e23}) ,
\begin{eqnarray*}
&&\sum_{a}\langle P,n|\hat{\Gamma}_{a}|P,n\rangle\hat{\Gamma}_{a}|P,n\rangle\\
&=&\frac{1}{2}\left[\langle P,n|A^{\nu}_{\mu}|P,n\rangle A^{\mu}_{\nu}-\frac{1}
{N}\sum_{\mu}A^{\mu}_{\mu}\right]|P,n\rangle=\lambda(P,n)|P,n\rangle\:,\\
\lambda(P,n)&=&\frac{1}{2}\left(\sum_{\mu}n^{2}_{\mu}-P^{2}/N\right)=
\frac{1}{2}\sum_{\mu}(n_{\mu}-P/N)^{2}  .
\end{eqnarray*}
The latter results in
\begin{equation}\label{e24}
C'_{2}T(g)|P,n\rangle=\lambda(P,n)T(g)|P,n\rangle\:.
\end{equation}

The eigen value $\lambda(P,n)$ attains the maximum for the highest weights,
for which $\sum_{\mu}n^{2}_{\mu}=P^{2}=\max$. The $CS \;|P,u\rangle$ belong to
the orbit of the highest weight $\{n\}=\{P,0,\ldots,0\}$. So we get:
\begin{equation}\label{e25}
C'_{2}|P,u\rangle=\frac{P^{2}(N-1)}{2N}|P,u\rangle\;
\end{equation}

One can introduce a dispersion of the square of the length of the isospin
vector \cite{ab12} ,
\begin{equation}\label{e26}
\Delta C_{2}=\langle\Psi|\sum_{a}\hat{\Gamma}^{2}_{a}|\Psi\rangle-\sum_{a}
\langle\Psi|\hat{\Gamma}_{a}|\Psi\rangle^{2}=\langle\Psi|C_{2}-C'_{2}|\Psi
\rangle\:.
\end{equation}
The dispersion serve as a measure of the uncertainty of the state
$|\Psi\rangle$.
Due to the properties of the operators $C_{2}$  and $C'_{2}$ , it is group
invariant and has the least value $P(N-1)/2$ for the orbits of highest weights,
paticularly for the $CS$ constructed, with respect to all the orbits of the
discrete basis (\ref{e3}) .
The relative dispersion of the square of the length of the isospin vector has
the value in the $CS$ :

\begin{equation}\label{e27}
\Delta C_{2}/C_{2}=\frac{N}{N+P}\:,
\end{equation}

\noindent and tends to zero with $h\rightarrow 0,\;h=\frac{1}{P}$ .That fact
indicates already, that the quantity $h$ plays here the role of the Planck
constant. In the Sect.5 this analogy is traced in more details.

\section {The $CS$ overlapping}

The overlapping of the CS can be evaluated in different ways. For instance,
using the completeness relation (20) and formulas (\ref{e19}) , (\ref{e4}) , we
 get:
\begin{eqnarray}
\langle P,u|P,v\rangle&=&\sum_{\{n\}}\langle P,u|P,n\rangle\langle P,n|P,v
\rangle\nonumber\\
&=&\sum_{\{n\}}\Psi_{P,\{n\}}(u)\Psi_{P,\{n\}}(\tilde{v})=\langle u,
\tilde{v}\rangle^{P}\:.\label{e28}
\end{eqnarray}

\noindent Comparing the result with eq. (\ref{e14}) , one can write
\begin{equation}
\langle P,u|P,v\rangle=\Psi_{P,\tilde{v}}(u)\;,\label{e29}
\end{equation}

\noindent what confirms once again, that the spaces $\Pi_{P}$ and $\tilde{\Pi}
_{P}$ are, in quantum mechanical sense, merely the spaces of abstract vectors
in the $CS$ representation.

Let $\Psi_{P}(u)$ be a vector $|\Psi\rangle$  in the $CS$ representation,
$\Psi
_{P}(u)=\langle P,u|\Psi\rangle$.Then the formula take place
\begin{equation}\label{e30}
\Psi_{P}(u)=\int\langle P,u|P,v\rangle\Psi_{P}(v){\rm d}\mu_{P}(\bar{v},v)\:.
\end{equation}

\noindent That means, the $CS$ overlapping playes the role of the $\delta$-
function in the $CS$ representation.

The modulus of the $CS$ overlapping (\ref{e28}) possesses of the properties:
\begin{eqnarray}
|\langle P,u|P,v\rangle|&<&1,\;\lim_{P\rightarrow\infty}|\langle P,u|P,v\rangle
=0,\   {\rm if }  u\neq v\:,\nonumber\\
|\langle P,u|P,v\rangle|&=&1,\  {\rm only,\  if }   u=v\:.\label{e31}
\end{eqnarray}

\noindent That follows from the Cauchy inequality for the scalar product (\ref
{e1}) ,  $|\langle u,\tilde{v}\rangle|\leq\sqrt{\langle u,\tilde{u}\rangle
\langle v,\tilde{v}\rangle}$,  and from the conditions on the parameters of the
 $CS$,   $\langle u,\tilde{u}\rangle=\langle v,\tilde{v}\rangle=1$.

One can introduce a function $s(u,v)$ of the coordinates of two points  of
the projective space $CP^{N-1}$ ,
\begin{equation}\label{e32}
s^{2}(u,v)=-\ln|\langle P,u|P,v\rangle|^{2}=-P\ln|\langle u,\tilde{v}\rangle|^
{2}\;.
\end{equation}
The properties of the modulus of the $CS$ overlapping (\ref{e31}) allows one to
interprete the function as  a symmetric. We remember, that a real and positive
symmetric obeys only two axioms of a distance: $s(u,v)=s(v,u)$ and $s(u,v)=0$,
if and only if $u=v$, exepting the triangle axiom. For the $CS$ of the
Heisenberg-Weyl group the function $s^{2}(u,v)=-\ln|\langle u|v\rangle|^{2}=
|u-v|^{2}$,  and gives real the square of the distance on the complex plane of
the $CS$ parameters. It turns out, that in case of consideration, the symmetric
 $s(u,v)$ generates the metric in the projective space $CP^{N-1}$ .To
demonstrate that, it is convenient to go over from the homogeneous coordinates
 $u_{\mu}$ ,  subjected to the
supplemental conditions (\ref{e18}) , to the local independent coordinates in
$CP^{N-1}$.
For instance, in the domain  where $u_{N}\neq 0$, we introduce the local
coordinates  $\alpha_{i},\:i=\overline{1,N-1}$,
\begin{eqnarray}
\alpha_{i}&=&u_{i}/u_{N}\:,\label{e33}\\
 u_{i}&=&\alpha_{i}u_{N},\;  u_{N}=\frac{\exp(-\frac{i}{N}\sum\arg\alpha_{k})}
{\sqrt{1+\sum|\alpha_{k}|^{2}}}\nonumber\:.
\end{eqnarray}

\noindent In the local coordinates (\ref{e33}) the symmetric (\ref{e32}) takes
the form
\begin{equation}\label{e34}
s^{2}(\alpha,\beta)=-P\ln\frac{\lambda(\alpha,\bar{\beta})\lambda(\beta,\bar{
\alpha})}{\lambda(\alpha,\bar{\alpha})\lambda(\beta,\bar{\beta})},
\end{equation}

\noindent where $\lambda(\alpha,\bar{\beta})=1+\sum_{i}\alpha_{i}\bar{\beta}_
{i}$ .

So, we are in position to calculate the square of the "distance" between
two infinitesimal close points $\alpha$ and $\alpha+{\rm d}\alpha$ .For the
${\rm d}s^{2}$, which is defined as the quadratic part of the decomposition of
$s^{2}(\alpha,\alpha+{\rm d}\alpha)$ in the powers of ${\rm d}\alpha$ , one
finds:
\begin{eqnarray}
{\rm d}s^{2}&=&g_{i\bar{k}}{\rm d}\alpha_{i}{\rm d}\bar{\alpha}_{k},\:
  g_{i\bar{k}}=P\lambda^{-2}(\alpha,\bar{\alpha})\left[\lambda(\alpha,\bar{
\alpha})\delta_{ik}-\bar{\alpha}_{i}\alpha_{k}\right]\:,\nonumber\\
g_{i\bar{k}}&=&\frac{\partial^{2}F}{\partial\alpha_{i}\partial\bar{\alpha}_{k}},
\;   F=P\ln\lambda(\alpha,\bar{\alpha})\:,\label{e35}\\
\det\|g_{i\bar{k}}\|&=&P^{N-1}\lambda^{-N}(\alpha,\bar{\alpha}),\
g^{\bar{k}i}=
\frac{1}{P}\lambda(\alpha,\bar{\alpha})(\delta_{ki}+\bar{\alpha}_{k}\alpha_{i})
\:.\nonumber
\end{eqnarray}
Now one can recognize in the expression (\ref{e35}) so called Fubini-Study
metric
 of
the complex projective space $CP^{N-1}$ with the constant holomorphic sectional
curvature $C=4/P$, \cite{ab16} . It follows from (\ref{e35}), we deal with
Kahlerian manifold. As known, a Kahlerian manifold is symplectic one and a
classical mechanics exists on it. The Poisson bracket has the form:
\begin{equation}\label{e36}
\{f,g\}=ig^{\bar{k}i}\left(\frac{\partial f}{\partial\alpha}_{i}\frac{\partial
 g}{\partial\bar{\alpha}_{k}}-\frac{\partial f}{\partial\bar{\alpha}_{k}}\frac
{\partial g}{\partial\alpha_{i}}\right)\:.
\end{equation}

\noindent In the next Sect. we show that the classical limit of the commutator
of the operators symbols, connected with the  $CS$, generates namely that
Poisson bracket.

\section {The classical limit}

One of the dignity of $CS$ is, they allow one to construct the operators
symbols
 in a simple way, i.e. a correspondance between operators and classical
functions on the phase space of a system. The reproduction of actions with
operators on the symbols language is, in fact, equivalent to the quantization
problem.That approach to the quantization was developed by Berezin \cite{ab6}.
Our aim has more restricted character, in that Sect. we are going to
investigate the conditions of the classical limit in terms of operators
symbols constructed by means of the $CS$.

Let us turn to the so called covariant symbol \cite{ab17} , which is, in fact,
the mean value of an operator $\hat{A}$ in the $CS$.
\begin{equation}\label{e37}
Q_{A}(u,\bar{u})=\langle P,u|\hat{A}|P,u\rangle\;.
\end{equation}

\noindent We also restrict ourself with operators which are some polynomial
functions on the generators, of power not more then some given one $M<P$. Such
 kind of operators can be written via the operators $a^{\dag}_{\mu}, a^{\nu}$,
using  (\ref{e10}) ,(\ref{e11}) , and be presented in the "normal" form,
\begin{equation}\label{e38}
\hat{A}=\sum_{K=0}^{M}A^{\mu_{1}\ldots\mu_{K}}_{\nu_{1}\ldots\nu_{K}}
a^{\dag}_{\mu_{1}}\ldots a^{\dag}_{\mu_{K}}a^{\nu_{1}}\ldots a^{\nu_{K}}
\:.
\end{equation}

\noindent It is easy to find the action of the operators $a^{\dag}_{\mu},  a^{
\nu}$ on the $CS$ and to calculate the symbols (\ref{e37}) ,
\begin{eqnarray}
a^{\dag}_{\mu}|P,u\rangle&=&\frac{1}{P+1}\frac{\partial}{\partial\tilde
{u}^{\mu}}|P+1,u\rangle,\;  a^{\mu}|P,u\rangle=P\tilde{u}^{\mu}|P-1,u\rangle\:,
\nonumber\\
\langle P,u|a^{\dag}_{\mu}&=&u_{\mu}\langle P-1,u|,\;  \langle P,u|a^{\mu}=
\frac{\partial}{\partial u_{\mu}}\langle P+1,u|\:.\label{e39}
\end{eqnarray}

\noindent So,
\begin{equation}\label{e40}
Q_{A}(u,\bar{u})=\sum_{K=0}^{M}\frac{P!}{(P-K)!}A^{\mu_{1}\ldots\mu_{K}}_{\nu_{1}
\ldots\nu_{K}}u_{\mu_{1}}\ldots u_{\mu_{K}}\bar{u}_{\nu_{1}}\ldots\bar{u}
_{\nu_{K}}\:.
\end{equation}

\noindent Obviously, there is a one-to-one correspondance between an operator
and its covariant symbol.

In the local independent variables $\alpha$ which were defined in (\ref{e33})
the covariant symbol has the form
\begin{equation}\label{e41}
Q_{A}(\alpha,\bar{\alpha})=\sum_{K=0}^{M}\frac{P!}{(P-K)!}\left(1+\sum_{i}^
{N-1}|\alpha_{i}|^{2}\right)^{-K}A^{\mu_{1}\ldots\mu_{K}}_{\nu_{1}\ldots
\nu_{K}}\alpha_{\mu_{1}}\ldots\alpha_{\mu_{K}}\bar{\alpha}_{\nu_{1}}\ldots
\bar{\alpha}_{\nu_{K}}\:,
\end{equation}

\noindent where the summation over the Greek indeces runs from $1$ to $N$ as
before, but one has to count $\alpha_{N}=1$ .

In manipulations it is convenient to deal with the nondiagonal symbols
\begin{eqnarray}
Q_{A}(u,\bar{v})&=&\frac{\langle P,u|\hat{A}|P,v\rangle}{\langle
P,u|P,v\rangle}
\label{e42}\\
&=&\sum_{K=0}^{M}\frac{P!}{(P-K)!}\left(\sum_{\lambda}^{N}u_{\lambda}\bar{v}_
{\lambda}\right)^{-K}\!\!A^{\mu_{1}\ldots\mu_{K}}_{\nu_{1}\ldots\nu_{K}}u_{\mu_
{1}}\ldots u_{\mu_{K}}\bar{v}_{\nu_{1}}\ldots\bar{v}_{\nu_{K}}\:,\nonumber\\
Q_{A}(\alpha,\bar{\beta})&=&\sum_{K=0}^{M}\frac{P!}{(P-K)!}\left(1+\sum
_{i}^{N-1}\alpha_{i}\bar{\beta}_{i}\right)^{-K}\!\!A^{\mu_{1}\ldots\mu_{K}}
_{\nu_{1}\ldots\nu_{K}}\alpha_{\mu_{1}}\ldots\alpha_{\mu_{K}}\bar{\beta}_{\nu
_{1}}\ldots\bar{\beta}_{\nu_{K}}\:,\nonumber\\
\alpha_{i}&=&u_{i}/u_{N},\;  \beta_{i}=v_{i}/v _{N},\;\alpha_{N}=\beta_{N}=1\:.
\nonumber
\end{eqnarray}

The symbols $Q_{A}(\alpha,\bar{\beta})$  are analytical functions on $\alpha$
and $\bar{\beta}$ separately and coincide with the covariant symbols
(\ref{e41})
 at $\beta\rightarrow\alpha$ . These symbols namely, but not $\langle P,\alpha|
A|P,\beta\rangle$ , are nondiagonal analytical continuation of the covariant
symbols.

Using the completeness relation and the eq.(\ref{e32}) , one can find for the
symbol of the product of two operators $\hat{A}_{1}$ and $\hat{A}_{2}$  :
\begin{equation}\label{e43}
Q_{A_{1}A_{2}}(u,\bar{u})=\int
Q_{A_{1}}(u,\bar{v})Q_{A_{2}}(v,\bar{u})e^{-s^{2}
(u,v)}{\rm d}\mu_{P}(\bar{v},v)\:.
\end{equation}

\noindent Because of $s^{2}(u,v)$ tends to infinity with $P\rightarrow\infty$,
if $u\neq v$ , and equals zero, if $u=v$, one can conclude, that in that limit
the domain $v\approx u$ gives only a contribution to the integral. Thus,
\begin{eqnarray}
\lim_{h\rightarrow 0}Q_{A_{1}A_{2}}(u,\bar{u})&=&Q_{A_{1}}(u,\bar{u})Q_{A_{2}}
(u,\bar{u})\int e^{-s^{2}(u,v)}{\rm d}\mu_{P}(\bar{v},v)\nonumber\\
&=&Q_{A_{1}}(u,\bar{u})Q_{A_{2}}(u,\bar{u})\:,\;  h=\frac{1}{P}\:.\label{e44}
\end{eqnarray}

\noindent The integral in (\ref{e44}) equals unity because of the definition
(\ref{e32})  and completeness relation.

If we define according to Beresin \cite{ab6,ab17}  the so called star
multiplication of symbols
\begin{equation}\label{e45}
Q_{A_{1}}\star Q_{A_{2}}=Q_{A_{1}A_{2}}\;,
\end{equation}

\noindent then we have for the covariant symbols
\begin{equation}\label{e46}
\lim_{h\rightarrow 0}Q_{A_{1}}\star Q_{A_{2}}=Q_{A_{1}}Q_{A_{2}}\;.
\end{equation}

\noindent That is the first demand of the classical limit. Thus, the quantity
$h$ plays the role of the Planck constant here, as we have already notice
before in Sect.3.

Now we are going to get the next term in the decomposition of the star
multiplication  (\ref{e45}) in powers of $h$. That is more appropriate to do
in the local independent coordinates $\alpha$ , because of the decomposition
includes the operation of differentiation with respect to coordinates. The
formula (\ref{e43}) in the local coordinates (\ref{e33}) takes the form
\begin{eqnarray}\label{e47}
Q_{A_{1}A_{2}}(\alpha,\bar{\alpha})&=&\int Q_{A_{1}}(\alpha,\bar{\beta})Q_{A_{2
}}
(\beta,\bar{\alpha})e^{-s^{2}(\alpha,\beta)}{\rm d}\mu_{P}(\bar{\beta},\beta)\:
,\\
{\rm d}\mu_{P}(\bar{\beta},\beta)&=&\frac{(P+N-1)!}{P!P^{N-1}}\det\|g_{l\bar
{m}}(\beta,\bar{\beta})\|\prod_{i=1}^{N-1}\frac{{\rm d}Re\beta_{i}{\rm d}Im
\beta_{i}}{\pi}\:,\nonumber
\end{eqnarray}
where ${\rm d}\mu_{P}(\bar{\beta},\beta)$ is proportional to the well known
 G-invariant measure on $CP^{N-1}$ (see eq.(35)).
Decomposing the integrand near the point $\beta=\alpha$, and going over to the
integration over $z=\beta-\alpha$, we get in the zero and first oder in power
$h$ :
\begin{eqnarray}
Q_{A_{1}A_{2}}(\alpha,\bar{\alpha})&=&Q_{A_{1}}(\alpha,\bar{\alpha})Q_{A_{2}}(
\alpha,\bar{\alpha})+\frac{\partial Q_{A_{1}}(\alpha,\bar{\alpha})}{\partial
\bar{\alpha}_{k}}\frac{\partial Q_{A_{2}}(\alpha,\bar{\alpha})}{\partial\alpha_
{i}}    \label{e48}\\
&\times&\det\|g_{l,\bar{m}}(\alpha,\bar{\alpha})\|\int\bar{z}_{k}z_{i}e^{-g_{a
\bar{b}}z_{a}\bar{z}_{b}}
\prod_{j=1}^{N-1}\frac{{\rm d}Re z_{j}{\rm d}Im z_{j}}{\pi}\nonumber \\
&=&Q_{A_{1}}(\alpha,\bar{\alpha})Q_{A_{2}}(\alpha,\bar{\alpha}) + g^{i\bar{k}}
\frac{\partial Q_{A_{1}}(\alpha,\bar{\alpha})}{\partial\bar{\alpha}_{k}}
\frac{\partial Q_{A_{2}}(\alpha,\bar{\alpha})}{\partial\alpha_{i}}\:,\nonumber
\end{eqnarray}
where the matrix $g^{i\bar{k}}$ was defined in (\ref{e35}) and is proportional
to $h$ . Taking into account the expression (\ref{e36})  for the Poisson
bracket in the projective space $CP^{N-1}$ , we get for the star commutator of
the symbols
\begin{equation}\label{e49}
Q_{A_{1}}\star Q_{A_{2}}-Q_{A_{2}}\star Q_{A_{1}}=i\{Q_{A_{1}},Q_{A_{2}}\}
+{\rm o}(h)\:.
\end{equation}
Thus, the second demand of the classical limit is satisfied.

\section{Conclusion}

One has to note, that the uniform discription of the $CS$ for all the groups
$SU(N)$,  proved to be possible due to the choice of the irreducible
representations
of the groups in the spaces of polynomials of a fixed power, so that the coset
space is parametrized by the homogeneous coordinates from $CP^{N-1}$. If one
constructs orbits, using the concrete structure of the matrices of the group
in the fundamental representation, as that was done in the majority of works
devoted to the $CS$ of the $SU(2)$ group, then the complications of the
generalization of the method are connected with the increasing of the
complicacy of the structure of the $SU(N)$ matrices with the growth of the
number $N$. The
representations in the spaces of polynomials of a power not greater then a
given one, used in \cite{ab7} for the $SU(2)$ group, lead at once to the
parametrization of the coset space by local coordinates from $CP^{N-1}$; due to
the nonsymmetrical form of the expressions in that case, the generalization to
any $SU(N)$ group does not appear to be obvious. Another approach to the
problem is possible in a Fock space, by means of representations of the Jordan-
Schwinger type. To construct explicitly  orbits here, it is necessary to
disentangle rather complicated operators of the representation of the group in
the Fock space. We can do that in the $SU(2)$ case, but the complicacy growth
with the number $N$ essentialy.

The $CS$ constructed are useful for the quasi-classical analysis of  quantum
equations of $SU(N)$ symmetrical gauge theories. With their help one can, for
 instance, derive the so colled Wong equation and find "quantum" corrections
to the equations  \cite{ab14} .

\end{document}